\documentclass[twocolumn,aps,showpacs]{revtex4}
\usepackage{color}
\usepackage{graphics}
\usepackage{epsfig}
\usepackage{amsmath}
\usepackage{amssymb}
\usepackage{amsthm}
\usepackage{amsmath,alltt}
\usepackage{graphicx}
\usepackage{amssymb}
\usepackage[latin1]{inputenc}
\usepackage[T1]{fontenc}
\usepackage{color}
\newcommand{\slsh}[1]{{\not \! #1}}

\newcommand{\bea}{\begin{eqnarray}}
\newcommand{\eea}{\end{eqnarray}}

\newcommand{\be}{\begin{equation}}
\newcommand{\ee}{\end{equation}}
\newcommand{\nn}{\nonumber}

\begin{document}

\title{Vacuum Polarization and Dynamical Chiral Symmetry Breaking: Phase Diagram of QED with Four-Fermion Contact Interaction}

\author{F. Akram$^1$, A. Bashir$^{2,3,4}$, L.X. Guti\'errez-Guerrero$^2$, B. Masud$^1$, J.~Rodr\'{\i}guez-Quintero$^5$, C. Calcaneo-Roldan$^6$, M.E. Tejeda-Yeomans$^6$ }
\affiliation{ $^1$ Centre for High Energy Physics, University of
the Punjab, Lahore, Pakistan. \\
$^2$Instituto de F\'isica y Matem\'aticas, Universidad Michoacana
de San Nicol\'as de Hidalgo, Edificio C-3, Ciudad Universitaria,
Morelia, Michoac\'an 58040, M\'exico.\\
$^3$Physics Division, Argonne National Laboratory, Argonne,
Illinois 60439, USA.  \\
$^4$Center for Nuclear Research, Department of Physics, Kent State
University, Kent Ohio 44242, USA. \\
$^5$ Dpto. F\'isica Aplicada, Fac. Ciencias Experimentales,
Universidad de Huelva, Huelva 21071, Spain. \\ $^6$Departamento de
F\'isica, Universidad de Sonora, Boulevard Luis Encinas J. y
Rosales, Colonia Centro, Hermosillo, Sonora 83000, M\'exico. }

\begin{abstract}

We study chiral symmetry breaking for fundamental charged fermions
coupled electromagnetically to photons with the inclusion of
four-fermion contact self-interaction term, characterized by
coupling strengths $\alpha$ and $\lambda$, respectively. We employ
multiplicatively renormalizable models for the photon dressing
function and the electron-photon vertex which minimally ensures
mass anomalous dimension $\gamma_m=1$. Vacuum polarization screens
the interaction strength. Consequently, the pattern of dynamical
mass generation for fermions is characterized by a critical number
of massless fermion flavors $N_f=N_f^c$ above which chiral
symmetry is restored. This effect is in diametrical opposition to
the existence of criticality for the minimum interaction
strengths, $\alpha_c$ and $\lambda_c$, necessary to break chiral
symmetry dynamically. The presence of virtual fermions dictates
the nature of phase transition. Miransky scaling laws for the
electromagnetic interaction strength $\alpha$ and the four-fermion
coupling $\lambda$, observed for quenched QED, are replaced by a
mean-field power law behavior corresponding to a second order
phase transition. These results are derived analytically by
employing the bifurcation analysis, and are later confirmed
numerically by solving the original non-linearized gap equation. A
three dimensional critical surface is drawn in the phase space of
$(\alpha, \lambda, N_f)$ to clearly depict the interplay of their
relative strengths to separate the two phases. We also compute the
$\beta$-functions ($\beta_{\alpha}$ and $\beta_{\lambda}$), and
observe that $\alpha_c$ and $\lambda_c$ are their respective
ultraviolet fixed points. The power law part of the momentum
dependence, describing the mass function, implies $\gamma_m =
1+s$, which reproduces the quenched limit trivially. We also
comment on the continuum limit and the triviality of QED.

\end{abstract}

\noindent 

\maketitle

\date{\today}

Since the works of Maskawa and Nakajima as well as the Kiev
group~\cite{Miransky:1976}, it is well known that quenched quantum
electrodynamics (QED) exhibits vacuum rearrangement, which
triggers chiral symmetry breaking when the interaction strength
$\alpha=e^2 / (4 \pi)$ exceeds a critical value $\alpha_c \sim 1$.
$\alpha_c$ was argued to be an ultraviolet stable fixed point
defining the continuum limit in supercritical QED. Although these
works were carried out for the bare vertex in the Landau gauge,
principle qualitative conclusions were later shown to be robust
even for the most general and sophisticated {\em ansätze} put
forward henceforth for an arbitrary value of the covariant gauge
parameter, see e.g.,~\cite{Pennington:1990, Gusynin:1994,
Saul:2011, Bermudez:2012}. Bardeen, Leung and
Love~\cite{Leung:1986} demonstrated that the composite operator
$\bar{\psi} \psi$ acquires large anomalous dimensions at
$\alpha=\alpha_c$. In fact, the mass anomalous dimension was shown
to be $\gamma_m=1$, leading to the fact that the four-fermion
interaction operator $(\bar{\psi} \psi)^2$ acquires the scaling
dimension of $d=2(3-\gamma_m)=4$ instead of 6, and becomes
renormalizable. This is an example of when an interaction which is
irrelevant in a certain region of phase space (perturbative) might
become relevant in another (non perturbative). Consequently, the
four-fermion contact interaction becomes marginal whose absence
cannot render QED a closed theory in the strong coupling domain.
Depending upon the non perturbative details of the fermion-boson
interaction, it is plausible to have $\gamma_m  >1$, implying $d <
4$, which would modify the status of the four-point operators from
marginal to relevant, see, e.g., the review
article~\cite{Bashir:2005}, and references therein. The upshot of
the argument is that the robustness of any conclusion about strong
QED can be guaranteed only if it is supplemented by these
perturbatively irrelevant operators. Quenched QED with the
inclusion of these additional operators has been studied
in~\cite{Gorbar:1991}.

Unquenching QED involves inclusion of fermion loops. It provides
screening and transforms the vacuum characteristics drastically,
changing the Miransky scaling law for the dynamically generated
mass $m \sim \Lambda \; {\rm Exp}[-\pi/\sqrt{\alpha/\alpha_c -
1}]$ to a mean field square-root behavior, i.e., $m \sim \Lambda
\; \sqrt{\alpha - \alpha_c}$,~\cite{Unquenched}. See
also~\cite{Roberts-Williams:1994} and references therein.
Employing a multiplicatively renormalizable photon propagator
proposed by Kizilersu and Pennington~\cite{Kizilersu:2009}, it has
recently been shown that large value of $N_f$ restores chiral
symmetry above a critical value $N_f^c$ and the corresponding
scaling law itself is a square-root: $m \sim \Lambda \;
\sqrt{N_f^c - N_f} $,~\cite{Calcaneo:2011}. However, these results
were demonstrated without incorporating four-fermion interactions.
In this article, we include this additional interaction and
establish the robustness of this result with the inclusion of all
the driving elements which influence chiral symmetry breaking,
namely, $\alpha$: the QED interaction strength, $\lambda$: the
coupling constant related to the four-fermion interactions and
$N_f$: the fermion flavors whose effect is diametrically opposed
to that of $\alpha$ and $\lambda$. We study the details of chiral
symmetry breaking in the vicinity of the phase change, mapping out
the phase space of all these relevant parameters and report the
results which survive as well as the ones that modify in different
regimes of this phase transition.

In section~\ref{Sec-Intro}, we introduce the framework of the
Schwinger-Dyson equations (SDEs), the notation as well as the
assumptions we employ for our analysis. Section~\ref{Sec-Analytic}
is dedicated to the analytic treatment of the gap equation in the
neighborhood of the critical plane which separates chirally
symmetric and asymmetric solutions. Next, in
section~\ref{Sec-Numerical}, we present the results of our
numerical analysis. The last section~\ref{Sec-Epilogue} summarizes
our findings and provides an outlook for future work.

\section{SDE for the Fermion Propagator} \label{Sec-Intro}

The starting point for our analysis is the SDE for the electron
propagator
 \bea
   S^{-1}(p) &=& S^{-1}_0(p) + i e^2 \int \frac{d^4k}{(2 \pi)^4} \gamma^{\mu} S(k) \Gamma^{\nu}(k,p)
 \Delta_{\mu \nu}(q)  \nn \\
 &-& i G_0 \, \int \frac{d^4k}{(2 \pi)^4} \; {\rm Tr} [S(k)]   \label{SDE-FP}\;,
 \eea
where $q=k-p$, $e$ is the electromagnetic coupling and $G_0$ is
the four-fermion coupling. We define the dimensionless
four-fermion coupling $\lambda$ as $\lambda / {\Lambda^2} = G_0 /
(4 \pi^2)$.
 $S^{-1}_0(p)=\slsh{p}$ is the inverse bare
propagator for massless electrons. We parameterize the full
propagator $S(p)$ in terms of the electron wave function
renormalization $F(p^2)$ and the mass function $M(p^2)$ as~$
   S(p) = {F(p^2)}/{(\slsh{p}-M(p^2))}.$
        $\Delta_{\mu \nu}(q)$  is the full photon propagator which can be
conveniently written as \bea \Delta_{\mu \nu}(q) =
-\frac{G(q^2)}{q^2} \left( g_{\mu \nu} - \frac{q_{\mu}
q_{\nu}}{q^2} \right) - \xi \frac{q_{\mu} q_{\nu}}{q^4}  \;, \eea
where $\xi$ is the covariant gauge parameter such that $\xi=0$
corresponds to the Landau gauge. $G(q^2)$ is the photon
renormalization function or the dressing function. The full
electron photon vertex is represented by $\Gamma^{\mu}(k,p)$. The
form of the full vertex is tightly constrained by various key
properties of the gauge theory,~\cite{Bashir:2005}, e.g.,
multiplicative renormalizability of the fermion and the gauge
boson propagators,~\cite{Pennington:1990, Bashir:1998,
Kizilersu:2009}, perturbation theory,~\cite{Bashir:2000}, the
requirements of gauge invariance/covariance,~\cite{Bermudez:2012,
Ward:1950, Bashir-1:1994, Bashir-2:1996, Landau:1956,
Bashir:LKF-1} and, of course, observed
phenomenology,~\cite{Craig:2009}. The most general decomposition
of this vertex in terms of its longitudinal and transverse
components is \bea \Gamma^{\mu}(k,p) = \sum_{i=1}^4 \lambda_i
(k,p) L_i^{\mu}(k,p) + \sum_{i=1}^8 \tau_i (k,p) T_i^{\mu}(k,p)
\;,  \label{FVertex} \eea where $L_1^{\mu}=\gamma^{\mu}$,
$L_2^{\mu}=(k+p)^{\mu}(\slsh{k}+\slsh{p})$,
$L_3^{\mu}=(k+p)^{\mu}$ and $L_4^{\mu}=\sigma^{\mu \nu}
(k+p)_{\nu}$, where $\sigma^{\mu \nu}= [\gamma^{\mu},
\gamma^{\nu}]/2$. The coefficients $\lambda_i$ are determined
through the Ward-Takahashi identity
 \bea
 (k-p)_{\mu} \Gamma^{\mu}(k,p) &=& S^{-1}(k)  - S^{-1}(p) \;,
 \eea
relating the electron propagator with the electron-photon
vertex,~\cite{Ball:1980}. Starting from the limiting form of this
identity, namely the Ward identity,
 \bea
   \frac{\partial S^{-1}(p)}{\partial p_{\mu}} &=&
   \Gamma^{\mu}(p,p) \;,
 \eea
 employing the most general form of the fermion propagator and
 then generalizing to arbitrarily different momenta, one obtains
\bea
   \lambda_1(k,p)&=&\frac{1}{2} \left[ \frac{1}{F(k^2)} + \frac{1}{F(p^2)} \right] \;, \nn \\
   \lambda_2(k,p)&=&\frac{1}{2}
\frac{1}{k^2-p^2} \left[ \frac{1}{F(k^2)} - \frac{1}{F(p^2)} \right] \;, \nn \\
   \lambda_3(k,p)&=&-\frac{1}{k^2-p^2} \left[ \frac{M(k^2)}{F(k^2)} - \frac{M(p^2)}{F(p^2)} \right] \;
\label{LVertex}
 \eea and $\lambda_4(k,p)=0$.

It has now been established that the choice of the transverse
vertex has observable consequences at the hadronic level, despite
the fact that the simple rainbow-ladder truncation is sufficient
to reproduce a large body of existing experimental data on
pseudoscalar and vector mesons such as their masses, charge radii,
decay constants and scattering lengths, as well as their form
factors and the valence quark distribution
functions,~\cite{Roberts-Maris:1997, Tandy-Maris:2000,
Tandy-Maris:2002, Ji-Maris:2001, Tandy-Maris:1999,
Tandy-Maris:2003, Maris:2002, Cotanch-Maris:2002,
Bashir-Guerrero-1:2010, Bashir-Guerrero-2:2010,
Tandy-Nguyen:2011}. For example, the conundrum of mass difference
between opposite parity states can only be explained through
corrections to the rainbow ladder
truncations,~\cite{Roberts-Chang:2009}, also see the
review,~\cite{ReviewCraig:2012}. In addition to the efforts
steered through the continuum studies, attempts have also been
initiated in lattice field theory to compute the transverse form
factors of the fermion-boson vertex in some simple kinematical
regimes,~\cite{Skullerud:2003,Skullerud:2007}. Extending these
efforts to the entire kinematical space of momenta $k^2, p^2$ and
$q^2$ is numerically challenging and it may require some time
before the results are made available. However, despite the fact
that the transverse vertex can have material effect on hadronic
properties and is crucial in maintaining key properties of a
quantum field theory, the qualitative behavior of the fermion mass
function itself is not significantly sensitive to its details.
Therefore, for our purpose, we shall restrict ourselves to the
simplest construction (Eq.~(8) of~\cite{Calcaneo:2011}) which, in
the quenched limit, renders the ultraviolet behavior of $M(p^2)$
to be
 \bea
      M(p^2) \sim (p^2)^{\gamma_m/2-1} \;, \label{anomalous}
 \eea
 with anomalous mass dimensions $\gamma_m=1$. This large value makes it
 mandatory to introduce four-point interactions to ensure self consistency.
 With this choice of the full vertex, we obtain, in the massless
 limit,
 \bea
     F(p^2) = \left( \frac{p^2}{\Lambda^2} \right)^{\nu} \;, \quad
 G(q^2) = \left( \frac{q^2}{\Lambda^2} \right)^s  \;,
\label{MRPropagators}
 \eea where $\nu= \alpha \xi / (4 \pi)$ and $s=
\alpha N_f / (3 \pi)$. Near criticality, where the generated
masses are small,
it is reasonable to assume that the power law solutions for the
propagators capture, at least qualitatively, correct description
of chiral symmetry breaking. We choose to study the resulting
equation for the mass function in the convenient Landau gauge.
Results for any other gauge can be derived by applying the
Landau-Khalatnikov-Fradkin transformations~\cite{Bashir:2009,
Bashir:LKF-1, Bashir:LKF-2} or using a vertex {\em ansatz} which
effectively incorporates gauge covariance properties in its
construction, e.g.,~\cite{Pennington:1990, Saul:2011,
Bermudez:2012}.

After taking the trace of~Eq.~(\ref{SDE-FP}), carrying out angular
integral and Wick rotating to Euclidean space, we obtain
\begin{widetext}
\begin{equation}
M(p^{2})=g(p^{2})\int\limits_{0}^{p^{2}}dk^{2}\frac{k^{2}}{p^{2}}\frac{%
M(k^{2})}{k^{2}+M^{2}(k^{2})}+\int\limits_{p^{2}}^{\Lambda ^{2}}dk^{2}\frac{%
M(k^{2})}{k^{2}+M^{2}(k^{2})}g(k^{2})+\frac{\lambda }{\Lambda ^{2}}%
\int\limits_{0}^{\Lambda ^{2}}dk^{2}\frac{k^{2}M(k^{2})}{k^{2}+M^{2}(k^{2})}%
,  \label{integral equation}
\end{equation}%
\end{widetext}

 \noindent where $g(p^{2})=s_{0} G(p^2)$,
$s_{0}=3\alpha /(4\pi )$ and
 $\Lambda $ is the
ultraviolet cutoff which regularizes the integrals. Note that we
have employed the simplifying assumption $G(q^2)=G(k^2)$ for $k^2
> p^2$ and $G(q^2)=G(p^2)$ for $p^2 > k^2$, which would allow for the
analytic treatment of the linearized  equation for the mass
function as detailed in the following section.

\section{Analytic Treatment}  \label{Sec-Analytic}

Before we venture into the computation of the mass function by
numerically solving the above non-linear integral equation, we
find it insightful to make analytical inroads. The differential
version of the gap equation (\ref{integral equation}) simplifies
in the neighborhood of the critical coupling $\alpha_c$; viz., the
coupling whereat $M(p^2) \neq 0 $ solution bifurcates away from
the $M(p^2) = 0 $ solution, which alone is possible in
perturbation theory. The behavior of the solution near the
bifurcation point may be investigated by performing functional
differentiation of the gap equation with respect to $M(p^2)$ and
evaluating the result at $M(p^2) = 0$. Practically, this amounts
to analyzing linearized form of the original gap equation [i.e.,
the equation obtained by eliminating all terms of quadratic or
higher order in $M(p^2)$]. \bea M(p^{2}) &=& \frac{g(p^{2})}{p^2}
\int\limits_{0}^{p^{2}}dk^{2} M(k^{2})
+\int\limits_{p^{2}}^{\Lambda ^{2}}dk^{2}\frac{%
M(k^{2})}{k^{2}}g(k^{2})   \nn \\
 &+&\frac{\lambda }{\Lambda ^{2}}%
\int\limits_{0}^{\Lambda ^{2}}dk^{2} \; {M(k^{2})}  \; .
\label{linearized}
 \eea

Note that the non-linearized version of this equation receives
negligible contribution from the region $k^2 \rightarrow 0$, while
this is not true for Eq.~(\ref{linearized}). This shortcoming is
readily remedied by introducing an infrared cutoff $m^2$ such that
$M(m)=m$. The resultant linearized gap
equation,~Eq.~(\ref{linearized}), can now be studied analytically
in the neighborhood of the critical plane on converting it into a
second order linear differential equation
 \begin{equation}
 x^{2}M^{\prime \prime }(x)+sxM^{\prime
}(x)+s_{0}(1-s)\frac{M(x)}{x^{s}}=0 \;, \label{linear DE}
 \end{equation}%
with two boundary conditions. Here we have used the convenient
substitution $x=\Lambda ^{2}/p^{2}$. Following are the infrared
and ultraviolet boundary conditions
\label{bc}
\begin{subequations}
\begin{eqnarray}
M^{\prime }(\Lambda ^{2}/m^{2}) &=&0 \;, \label{infra bc}
\\
M(1) &=&\left( 1+\frac{\lambda }{s_{0}}\right) \frac{M^{\prime
}(1)}{1-s} \;. \label{ultra bc}
\end{eqnarray}%
\end{subequations}
Note that the four-fermion coupling only affects the ultraviolet
boundary condition. The differential equation itself and the
infrared boundary condition do not have any direct dependence on
it. It is what we expect intuitively. four-fermion
Nambu--Jona-Lasinio (NJL) type term only generates a constant mass
term. Therefore, it effectively serves as a cut-off dependent bare
mass and can hence only influence the dynamics through the
ultraviolet boundary condition.

If we now apply the Lommel transformations: $z=Bx^{\gamma }$ and
$W(x)=x^{-a}M(x)$, the linearized equation can be converted into
the following Bessel differential equation
\begin{equation}
z^{2}W^{\prime \prime }(z)+zW^{\prime }(z)+(z^{2}-A^{2})W(z)=0 \;,
\label{bessel DE}
\end{equation}%
where, $\gamma =-s/2,$ $a=(1-s)/2,$ $A=(1-s)/s,$ and $B=\sqrt{%
3\alpha (1-s)/(\pi s^{2})}$, where $s<1$ . The boundary conditions
in terms of the function $W(z)$ are given by
\begin{subequations}
\label{bc of w}
\begin{eqnarray}
\hspace{-1cm} a W(z)+\gamma zW^{\prime }(z)|_{z=B(m/\Lambda )^{s}}
&=&0 \; , \\ \nn \\\hspace{-1cm} a \left( s_0 - \lambda \right)
W(z) - \gamma z \left(s_0+ \lambda  \right)  W^{\prime }(z)|_{z=B}
&=& 0 \; .
\end{eqnarray}%
The general solution of the second order differential equation
(\ref{bessel DE}) is
\end{subequations}
\begin{equation}
W(z)=c_{1}J_{A}(z)+c_{2}Y_{A}(z) \;,  \label{sol}
\end{equation}%
where $J_{A}(z)$ and $Y_{A}(z)$ are the Bessel functions of the
first and the second kind, respectively. The power law part of
momentum dependence of the mass function $M(x)=x^{a} W(x)$ is
neatly separated out into the factor $x^a$, implying
$\gamma_m=1+s$. Note that $s=1$ corresponds to a {\em momentum
independent} photon propagator which implies $\gamma_m=2$.
Consequently, Eq.~(\ref{anomalous}) implies that it corresponds to
a momentum independent mass function, a result which is readily
and analytically confirmed from the resultant simple integral
equation. This is a well known behavior of a contact interaction
model of the NJL type. Moreover, the quenched limit of
$\gamma_m=1$ is also reproduced trivially.

\begin{figure}[!h]
\begin{center}
\includegraphics[angle=0,width=0.45\textwidth]{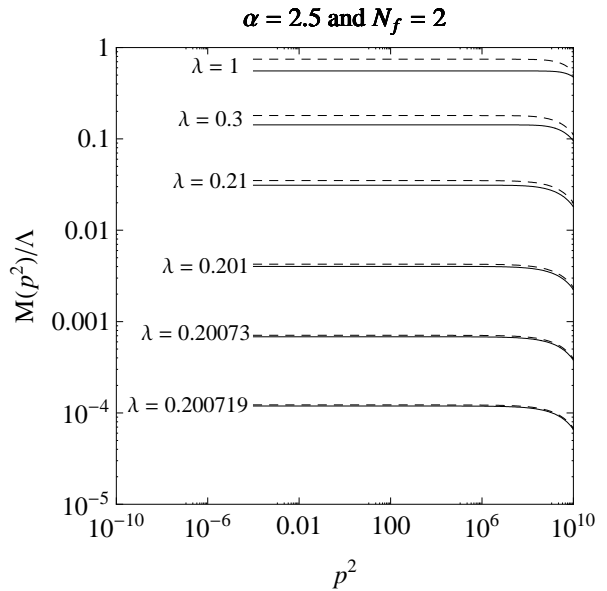}
\end{center}
\caption{The mass functions for different values of $\lambda$ at
fixed $\alpha = 2.5$ and $N_{f} = 2$. Its increasing sensitivity
to the variation in lambda helps locate the critical strength
$\lambda_c$. Dashed and solid curves represent mass functions
with~(vacuum polarization of Eq.~(\ref{Prop-feedback})) and
without feedback~(vacuum polarization of
Eq.~(\ref{MRPropagators})) from the gap equation, respectively.}
\label{fig1}
\end{figure}

\begin{figure}[!h]
\begin{center}
\includegraphics[angle=0,width=0.45\textwidth]{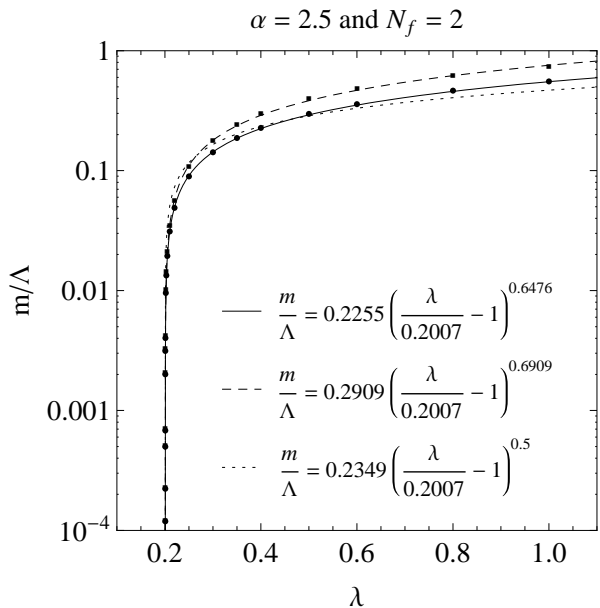}
\end{center}
\caption{The scaling law for four-fermion coupling $\lambda$.
Dots, solid and dashed lines represent numerical results, fit to
the numerical data with a power law and analytically predicted
square-root scaling law, respectively, at $\alpha = 2.5$ and
$N_{f} = 2$. The mean field behavior of the chiral phase
transition is evident. Dashed and solid curves represent mass
functions with~(vacuum polarization of Eq.~(\ref{Prop-feedback}))
and without feedback~(vacuum polarization of
Eq.~(\ref{MRPropagators})) from the gap equation, respectively.}
\label{fig2}
\end{figure}

For the homogenous boundary conditions of Eqs.~(\ref%
{bc of w}), the non-trivial chirally asymmetric solution of the
gap equation exists if the following condition holds~:
\begin{widetext}
\begin{equation}
\left[ \frac{2aJ_{A}(z)+\gamma z\left( J_{A-1}(z)-J_{A+1}(z)\right) }{%
2aY_{A}(z)+\gamma z\left( Y_{A-1}(z)-Y_{A+1}(z)\right) }\right] _{z=B\left(
m/\Lambda \right) ^{s}}=\frac{\left( 1+\lambda /s_{0}\right) \gamma B\left(
J_{A-1}(B)-J_{A+1}(B)\right) -(1-s)(1-\lambda /s_{0})J_{A}(B)}{\left(
1+\lambda /s_{0}\right) \gamma B\left( Y_{A-1}(B)-Y_{A+1}(B)\right)
-(1-s)(1-\lambda /s_{0})Y_{A}(B)}.
\end{equation}%
In the limit of $\Lambda \rightarrow \infty $, we obtain the
following result for the dynamically generated mass $m$~:
\begin{eqnarray}
\frac{m^{2}}{\Lambda ^{2}} \equiv f\left( \alpha,N_{f},\lambda
\right)   =\left[ \frac{2}{B}\right]^{\frac{2}{s}}\frac{\Gamma
(A)\Gamma (A+2)2a}{\pi \gamma } \left[ \frac{\left( 1+\lambda
/s_{0}\right) \gamma B\left( J_{A-1}(B)-J_{A+1}(B)\right)
-(1-s)(1-\lambda /s_{0})J_{A}(B)}{\left( 1+\lambda /s_{0}\right)
\gamma B\left( Y_{A-1}(B)-Y_{A+1}(B)\right) -(1-s)(1-\lambda
/s_{0})Y_{A}(B)}\right] .  \label{mass1}
\end{eqnarray}%
\end{widetext}
Carrying out the Taylor expansion near the critical point, we find
the following scaling laws~:
 \bea
 \frac{m}{\Lambda } &=& A_1 \left( \alpha -\alpha
_{c}\right)^{1/2} \;, \; \qquad N_f, \lambda \; \; {\rm fixed}  \\
 \frac{m}{\Lambda } &=& A_2 \left( \lambda -\lambda
_{c}\right)^{1/2} \;, \; \qquad N_f, \alpha \; \; {\rm fixed}  \\
 \frac{m}{\Lambda } &=& A_3 \left( N_f^c - N_f \right)^{1/2} \,. \, \quad \lambda, \alpha \; \; {\rm fixed}
\label{scaling law}
 \eea
The momentum dependence of the mass function (based upon the
numerical calculation discussed in the next section) and the
scaling laws for $\lambda, \alpha$ and $N_f$ have been plotted in
Figs.~(\ref{fig1}-\ref{fig5}).

The critical values of the parameters $\alpha $, $N_{f}$, and
$\lambda $ define a surface in the 3D phase-space of these
parameters. The mass function is zero (non-zero) below (above) the
critical surface, which corresponds to restored (broken) chiral
symmetry. The analytic expression for the critical surface can be
obtained by setting $m/\Lambda =0$ in equation (\ref{mass1}). The
resultant equation, which can be solved for $\lambda $, is given
by
\begin{eqnarray}
\lambda = \hspace{-1mm} -s_{0}\frac{\left[ \gamma B\left(
J_{A-1}(B)-J_{A+1}(B)\right) -(1-s)J_{A}(B)\right] }{\left[ \gamma
B\left( J_{A-1}(B)-J_{A+1}(B)\right) +(1-s)J_{A}(B)\right] }.
\label{criticality}
\end{eqnarray}
In order to obtain a finite mass for the charged fermion in the
limit of $\Lambda \rightarrow \infty$, one requires charge
renormalization. Therefore, in this limit, we impose $ \alpha
(\Lambda) = \alpha_c +{m^2}/{(A^2 \Lambda^2)}$. One can thus
readily obtain the corresponding $\beta$-function~:
 \bea
 \beta_{\alpha}  &=&  \Lambda \; \frac{ \partial \alpha}{\partial \Lambda}
 \Big{|}_{\lambda,N_f} = -2 (\alpha - \alpha_c) \;.
 \eea
Therefore, $\alpha_c$ is the ultraviolet fixed point of
$\beta_{\alpha}$, as has been observed in~\cite{Unquenched}.
Identical presence of the fixed point for $\beta_{\lambda}$ for
$\lambda_c$ is readily observed~:
 \bea
 \beta_{\lambda}  &=&  \Lambda \; \frac{ \partial \lambda}{\partial \Lambda}
 \Big{|}_{\alpha,N_f} = -2 (\lambda - \lambda_c) \;.
 \eea
The analytical results of this section can be confirmed and made
precise through a numerical study of the non-linearized gap
equation~(\ref{integral equation}). This analysis is presented in
the next section.

\section{Numerical Results}   ~\label{Sec-Numerical}

\begin{figure}[!h]
\begin{center}
\includegraphics[angle=0,width=0.45\textwidth]{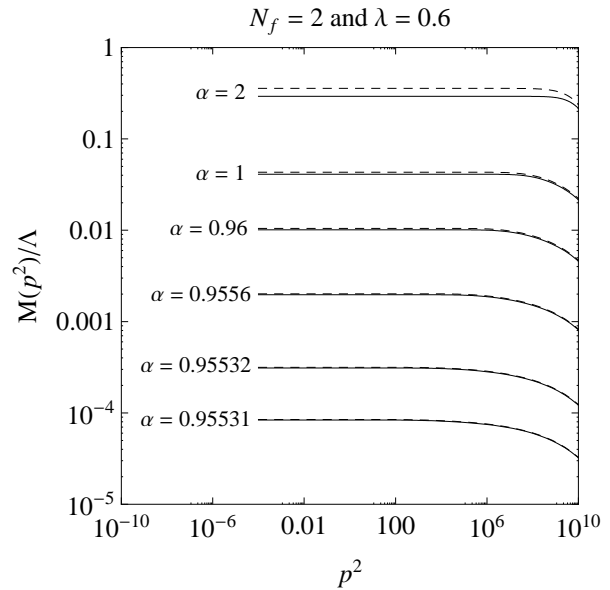}
\end{center}
\caption{ The mass function $M(p^2)$ for varying values of the
electromagnetic coupling $\alpha$ for fixed values of massless
fermion flavors $N_f=2$ and the four-fermion coupling
$\lambda=0.6$. The objective is to hunt $\alpha_c$ and determine
the nature of the phase transition. Dashed and solid curves
represent mass functions with~(vacuum polarization of
Eq.~(\ref{Prop-feedback})) and without feedback~(vacuum
polarization of Eq.~(\ref{MRPropagators})) from the gap equation,
respectively.} \label{fig3}
\end{figure}

\begin{figure}[!h]
\begin{center}
\includegraphics[angle=0,width=0.45\textwidth]{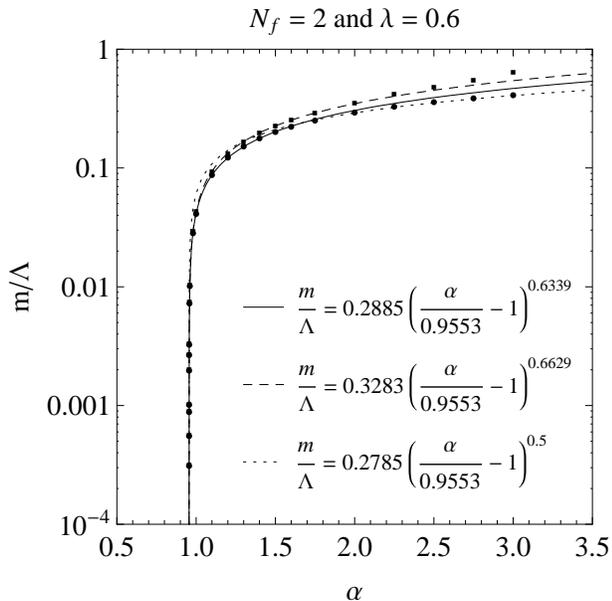}
\end{center}
\caption{The scaling law for the coupling $\alpha$, investigated
through the behavior of the mass function near
criticality,~Fig.~\ref{fig3}. Dots, solid and dashed lines
represent numerical results, fit of the numerical data to the
power law and analytically predicted square-root scaling law,
respectively at $N_{f} = 2$ and $\lambda = 0.6$. Dashed and solid
curves represent mass functions with~(vacuum polarization of
Eq.~(\ref{Prop-feedback})) and without feedback~(vacuum
polarization of Eq.~(\ref{MRPropagators})) from the gap equation,
respectively.} \label{fig4}
\end{figure}

\begin{figure}[!h]
\begin{center}
\includegraphics[angle=0,width=0.45\textwidth]{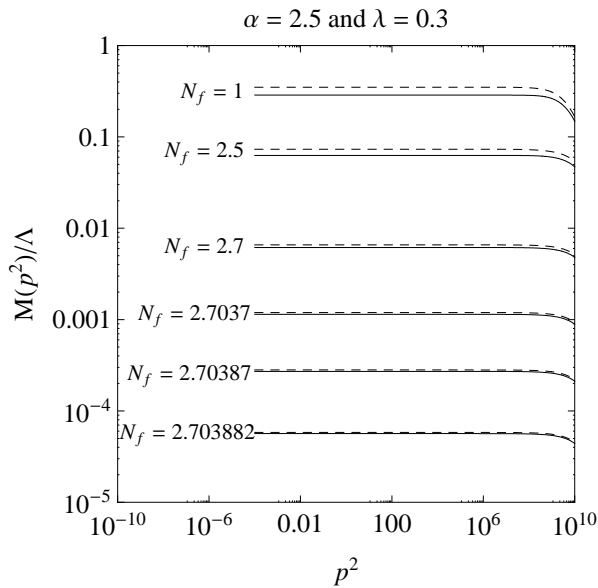}
\end{center}
\caption{The mass function $M(p^2)$ for increasing number of
massless fermion flavors diminishes because the interaction gets
screened. Chiral symmetry is restored above a certain $N_f^c$,
which depends upon the values of $\alpha$ and $\lambda$. Dashed
and solid curves represent mass functions with~(vacuum
polarization of Eq.~(\ref{Prop-feedback})) and without
feedback~(vacuum polarization of Eq.~(\ref{MRPropagators})) from
the gap equation, respectively.} \label{fig5}
\end{figure}

\begin{figure}[!h]
\begin{center}
\includegraphics[angle=0,width=0.45\textwidth]{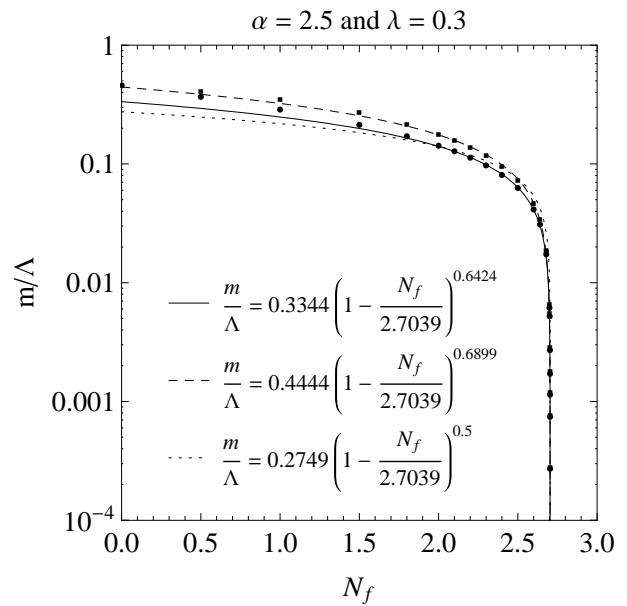}
\end{center}
\caption{ The scaling law for $N_{f}$. Dots, solid and dashed
lines represent numerical results, fit of the numerical data to
the power law and analytically predicted square-root scaling law,
respectively, at $\alpha = 2.5$ and $\lambda = 0.3$. Thus the
nature of this transition is independent of the inclusion of
four-fermion interaction term. Dashed and solid curves represent
mass functions with~(vacuum polarization of
Eq.~(\ref{Prop-feedback})) and without feedback~(vacuum
polarization of Eq.~(\ref{MRPropagators})) from the gap equation,
respectively.} \label{fig6}
\end{figure}

In order to compare and confirm the above analytical results,
based on the linearized approximation, we solve the original
non-linear integral equation (\ref{integral equation}) numerically
for varying $N_{f}$, $\alpha $, and $\lambda$. Depicted in Fig. 1
is the fermion mass functions for different values of $\lambda$
for $\alpha =2.5$ and \thinspace $N_{f}=2$. The closer we get to
the critical value $\lambda_c$, the more drastically pronounced is
the drop in the mass function, indicating the approaching
onslaught of the phase transition. The scaling law is explored in
Fig. 2, where the variation of $m/\Lambda \equiv
M(p^{2}=0)/\Lambda $ with $\lambda $ is plotted at the fixed
values of $\alpha $ and $N_{f}$. The fit of the complete numerical
data shows that the power of the scaling law is slightly different
from 0.5. This is expected as the mean field scaling behavior
(\ref{scaling law}) captures the correct physics in the immediate
vicinity of the critical point, only where the linearized version
of the equation becomes exact. In the same figure, we also
superimpose the analytically derived square-root scaling law
which, as expected, sits exactly atop the numerical findings in
the immediate vicinity of the critical point. In Figs.~\ref{fig3}
and \ref{fig4}, we show the variation of the mass section and the
corresponding scaling law as a function of $\alpha$ at fixed
values of $N_{f}=2$ and $\lambda =0.6$. These results again
confirm the validity of the square-root dependence of the
dynamically generated mass on the electromagnetic coupling. In
Figs.~\ref{fig5} and~\ref{fig6}, we plot the mass function in the
presence of increasing types of virtual fermion--anti-fermion
pairs and the resulting scaling law as a function of $N_{f}$,
respectively, for $\alpha =2.5$ and $\lambda =0.3$. These results
establish the robustness of the conclusions presented in the
reference~\cite{Calcaneo:2011} on the inclusion of the four-point
contact interaction term.

\begin{figure}[!h]
\begin{center}
\includegraphics[angle=0,width=0.45\textwidth]{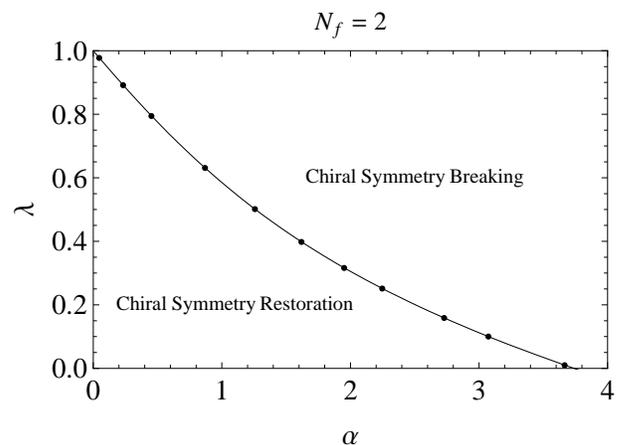}
\end{center}
\caption{Critical curve in $\lambda - \alpha $ plane at $N_{f}=2$.
Dots and solid curve represent the numerical results and
analytical findings, respectively. For a fixed $N_f$, chiral
symmetry breaking phase is achieved when the combined strength of
$\alpha$ and $\lambda$ lies above the criticality curve, dictated
by Eq.~(\ref{criticality}). The curve is indistinguishable,
independently of the photon propagator employed
[i.e.,~Eq.~(\ref{MRPropagators}) or~Eq.~(\ref{Prop-feedback})].
The same is true for Figs. \ref{fig8}, \ref{fig9} and
\ref{fig10}.} \label{fig7}
\end{figure}

\begin{figure}[!h]
\begin{center}
\includegraphics[angle=0,width=0.45\textwidth]{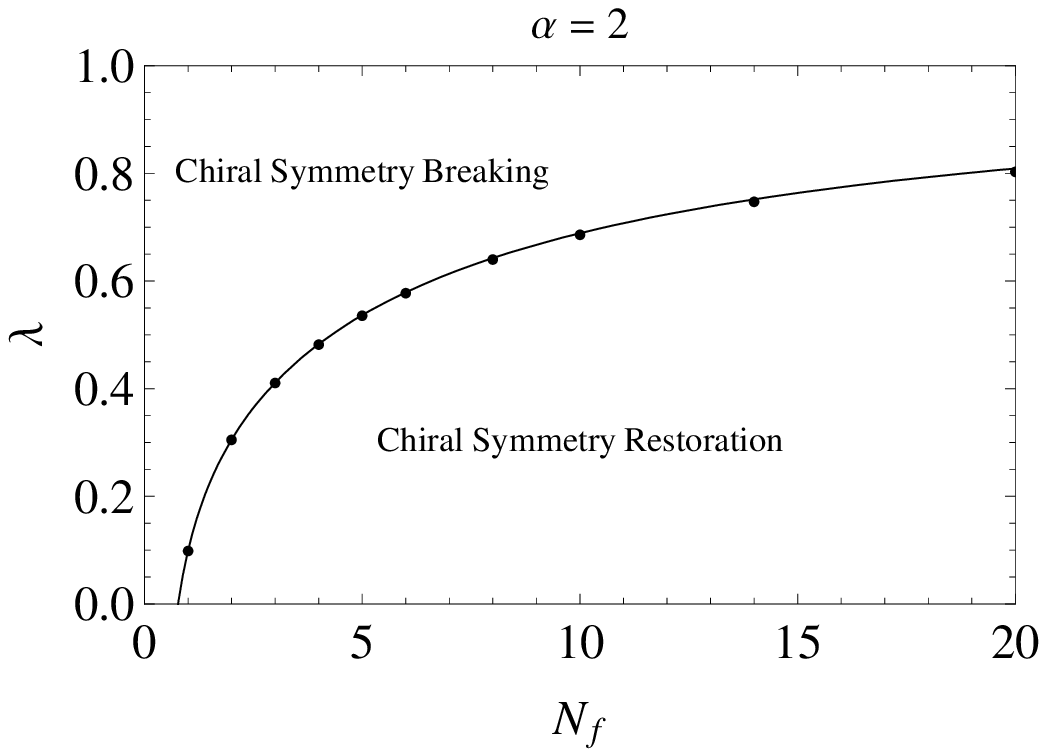}
\end{center}
\caption{Critical curve in $\lambda - N_{f} $ plane at $\alpha=2$.
Dots and solid curve represent the numerical results and
analytical findings, respectively. It is clear that the
bifurcation analysis provides an exact analysis of the non-linear
equation at criticality.} \label{fig8}
\end{figure}

\begin{figure}[!h]
\begin{center}
\includegraphics[angle=0,width=0.45\textwidth]{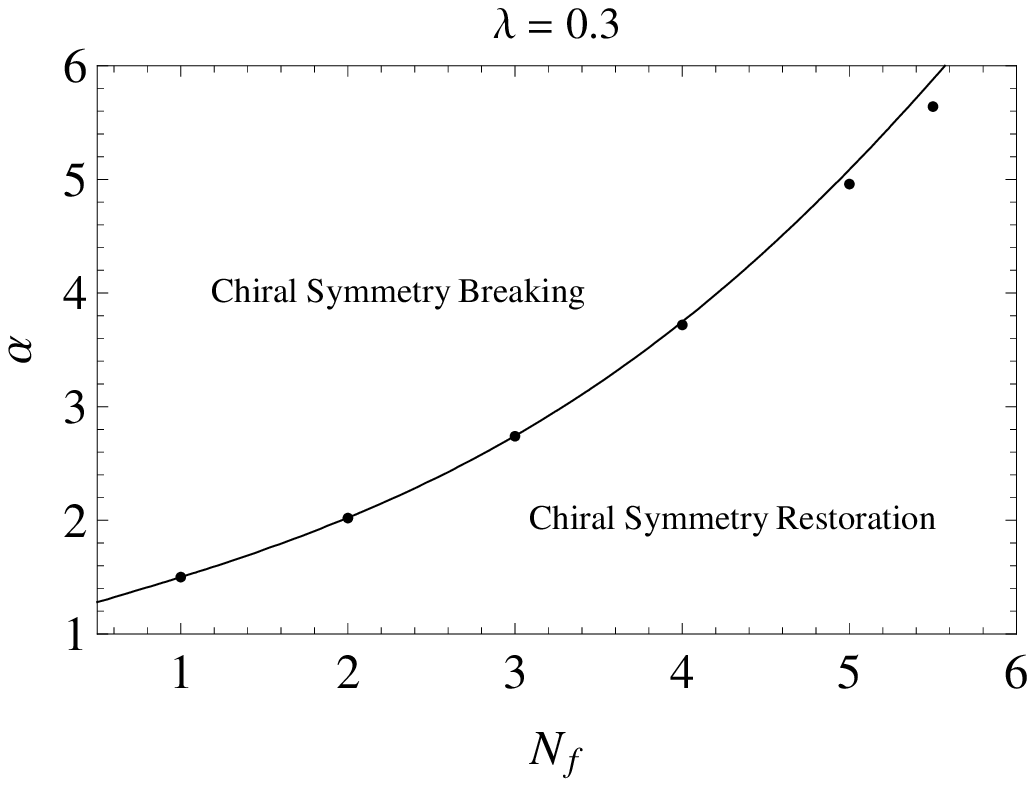}
\end{center}
\caption{Critical curve in $N_f - \alpha $ plane at $\lambda=0.3$.
As in the other curves, dots are numerical solutions whereas the
solid line is the outcome of the bifurcation analysis.}
\label{fig9}
\end{figure}

Shown in Fig.~\ref{fig7} is the critical curve in $\lambda -\alpha $ plane at $%
N_{f}=2$. The dynamical mass ceases to exist for the values of
$\lambda $ and $\alpha $ below the curve. The dots in this figure
represent the points obtained by numerically solving the
non-linear integral equation of the mass function and the solid
curve represents the analytical result which corresponds to
expression (\ref{criticality}). This criticality should be
considered as an extension of its quenched QED counterpart
obtained in~\cite{Yamawaki:1989}. For the sake of completeness, in
Figs.~\ref{fig8} and~\ref{fig9}, we present the critical curves in
$\lambda -N_{f}$ and $\alpha-N_f$ planes for fixed values of
$\alpha =2$ and $\lambda=0.3$, respectively. These figures show
that the analytical results agree with the numerical findings with
a very good accuracy. Fig.~(\ref{fig9}) gives a quantitative
picture of how the growing number of fermion flavors requires
stronger electromagnetic coupling. The relation is not linear. The
screening effect exhausts the strength of the interaction faster
with increasing $N_f$.

\begin{figure}[!h]
\begin{center}
\includegraphics[angle=0,width=0.45\textwidth]{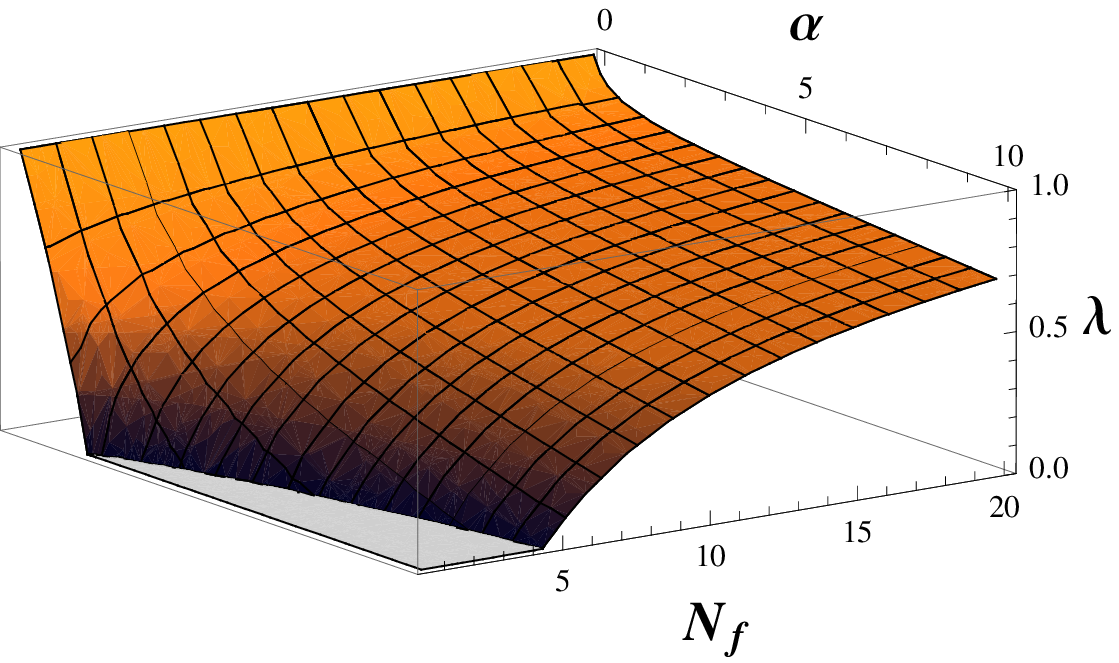}
\end{center}
\caption{A three dimensional view of the criticality surface. It
corresponds to Eq.~(\ref{criticality}) as well as the numerical
analysis of the non-linearized Eq.~(\ref{integral equation}). The
region below the surface represents chirally symmetric phase.}
\label{fig10}
\end{figure}

Finally, in Fig.~\ref{fig10}, we present the full critical surface
in the phase space of $N_{f}$, $\alpha $, and $\lambda $. In the
limit of $N_f \rightarrow 0 $, Miransky scaling law is reproduced.
The results presented in this paper are qualitatively robust if,
instead of the multiplicatively renormalizable photon propagator
of~Eq.~(\ref{MRPropagators}), we employ any of the following
models~:

\begin{itemize}

\item

One loop perturbative photon propagator, as employed
in~\cite{Unquenched}~:
 \bea
 G(q^2) &=& 1 + \frac{\alpha N_f}{3 \pi} \; {\rm ln} \left(  \frac{q^2}{\Lambda^2} \right)
 \;.
 \eea

\item

A photon propagator which receives feedback from the gap equation,
i.e.,
 \bea
  G(q^2) &=& [  \left( q^2 + M^2(0) \right)/ \Lambda^2]^s \;,
  \label{Prop-feedback}
 \eea
which is also multiplicatively renormalizable. For a comparison,
we have also displayed numerical results for this latter choice in
all the relevant figures. As we had anticipated, near criticality,
results are practically indistinguishable from the ones obtained
from using the model of~Eq.~(\ref{MRPropagators}).

\end{itemize}

Note that away from criticality, a complete self consistent
coupled solution of the photon and the fermion propagator will be
required. However, finiteness of the dynamically generated mass
for $\Lambda \rightarrow \infty$ forces non perturbative QED to be
consistently defined only for those values of $\alpha$ and
$\lambda$ which lie on the critical surface. This is a simple
corollary of the argument laid out in~\cite{Leung:1986}.
 Note that as we employ a model for the vacuum polarization, the running
coupling is not our prediction. Following
Rakow,~\cite{Rakow:1991}, if we define the renormalization at
$q^2=0$ rather than on the mass shell, we get
 \bea
 \alpha_R(0) = \alpha \, F_R^2(0) \, G_R(0) = \alpha \, G_R(0) \;,
 \eea
 because $F_R(0)=1$ for us.
 Therefore, our model conforms to $\alpha_R(0) \rightarrow 0$
 in accordance with the lattice computation of unquenched
 QED,~\cite{Rakow:1991}. As we have argued before, when $\Lambda \rightarrow
 \infty$, $\alpha \rightarrow \alpha_c $.
 Thus $\alpha_R(0)= \alpha_c G_R(0) \rightarrow 0$, which is associated with the
 triviality of QED in~\cite{Rakow:1991}. We use this same model for the vacuum polarization
 even in the presence of the perturbatively irrelevant four-fermion interaction terms. This means
 that on the phase boundary, the renormalised coupling is zero even in the presence of
 the four-fermion interactions. This is in accordance with the argument
 presented in~\cite{Roberts-Williams:1994}.
 However, note that for the
 practical solution of the gap equation, the photon propagator or
 the running coupling below $q^2 = \kappa^2 = M^2(\kappa^2)$ has no
 bearing on chiral symmetry breaking solution.

 \noindent
 We now recall that the Bethe-Salpeter
 equation gives us an approximate relation
 between the integral over the mass function and the "decay
 constant of the pion ($f$)" given by Eq. (4.42)
 of~\cite{Roberts-Williams:1994}
 \bea
 f^2 = \int^{\Lambda} dp^2 p^2 M(p^2) \; \frac{ \left[ M(p^2)- p^2
 M^{\prime}(p^2)/2\right]
 }{[p^2 + M^2(p^2)]^2} \;.
 \eea
 We numerically evaluate $f$ for varying values of $s$ and find that
 $f \rightarrow \infty$ in the limit when the ultraviolet regulator
 $\Lambda \rightarrow \infty$, a result which suggests that the continuum limit
 is the one of noninteracting bosons, in agreement with earlier
 findings~\cite{Unquenched, Kondo:1990}.

\section{Epilogue} \label{Sec-Epilogue}

We have studied chiral symmetry breaking for fundamental fermions
interacting electromagnetically with photons and self interacting
through perturbatively irrelevant four-fermion contact interaction
which is required to render QED a closed theory in its strongly
coupled regime, where, this additional interaction becomes
marginal and, perhaps, even relevant. We have used
multiplicatively renormalizable models for the photon propagator
(with and without the feedback from the gap equation) which, we
argue, should capture the qualitative physics correctly in the
vicinity of the critical surface in the phase space of $(\alpha,
\lambda, N_f)$, marking the onslaught of chiral symmetry breaking
(restoration). The presence of virtual fermion--anti-fermion pairs
changes the nature of the phase transitions along the $\alpha$ and
$\lambda$-axes. The Miransky scaling law softens down to a square
root mean field scaling behavior as a function of all the three
parameters~$\alpha$, $\lambda$ and $N_f$,~(\ref{scaling law}).
Study of the mass anomalous dimensions for QED with a model vacuum
polarization reveals how, quantitatively, the momentum dependence
of the photon propagator, {\em i.e.}, $(p^2)^{s-1}$, filters into
the momentum dependence of the fermion mass function, namely,
$(p^2)^{(s-1)/2}$, through the gap equation. We believe that our
analysis can and should be extended to the study of QCD through
its SDEs. Situation is ripe for the application of this line of
approach and technology to QCD, where we are finally having the
first glimpses of the flavor dependence of the gluon propagator in
the infrared region,~\cite{gluon}. This is for future.\\

\noindent {\bf Acknowledgments} AB acknowledges CIC (UMICH) and
CONACyT Grant Nos. 4.10, 46614-I, 128534, 94527 (Estancia de
Consolidaci\'on) and the US Department of Energy, Office of
Nuclear Physics, Contract No. DE-AC02-06CH11357. We thank V.
Gusynin, A. K{\i}z{\i}lers$\ddot{\rm u}$ and C.D. Roberts for
useful discussions.


\begin{thebibliography}{55}


\bibitem{Miransky:1976} T. Maskawa and H. Nakajima, Prog. Theor. Phys. {\bf 52}, 1326
(1974); P.I. Fomin and V.A. Miransky, Phys. Lett. B {\bf 64}, 166
(1976); P.I. Fomin, V.P. Gusynin and V.A. Miransky, Phys. Lett. B
{\bf 78}, 136 (1978).

\bibitem{Pennington:1990} D.C. Curtis and M.R. Pennington, Phys. Rev. D {\bf 42}, 4165
(1990);

\bibitem{Gusynin:1994} D. Atkinson, J.C.R. Bloch, V.P. Gusynin, M.R. Pennington and
M. Reenders, Phys. Lett. B {\bf 329}, 117 (1994).

\bibitem{Saul:2011} A. Bashir, A. Raya, S. S\'anchez-Madrigal,
Phys. Rev. D {\bf 84}, 036013 (2011).

\bibitem{Bermudez:2012} A. Bashir, R. Bermudez, L. Chang, C.D. Roberts, Phys. Rev. C {\bf
85}, 045205 (2012).

\bibitem{Leung:1986} W.A. Bardeen, C.N. Leung and S.T. Love, Phys. Rev. Lett. {\bf 56}, 1230
(1986); C.N. Leung, S.T. Love and W.A. Bardeen, Nucl. Phys. B {\bf
273}, 649 (1986).

\bibitem{Bashir:2005} A. Bashir and A. Raya, {\em Trends in Boson Research}, edited by A.V. Ling,
Nova Science Publishers, Inc. N. Y., ISBN: 1-59454-521-9 (2005)

\bibitem{Gorbar:1991} E.V. Gorbar, E.Sausedo, Ukr. Phys. J. {\bf 36}, 1025
(1991); Manuel Reenders (Groningen U.), Ph.D. Thesis, {\em
"Dynamical symmetry breaking in the gauged Nambu-Jona-Lasinio
model."} e-Print: hep-th/9906034 [hep-th] (1999).

\bibitem{Salmhofer:1991} M.Salmhofer and E.Seiler, Commun. Math. Phys. {\bf 139},
395 (1991).

\bibitem{Unquenched} K. Kondo, Y. Kikukawa and H. Mino, Phys.
Lett. B {\bf 220}, 270 (1989); V.P. Gusynin, Mod. Phys. Lett. A
{\bf 5}, 133 (1990).

\bibitem{Roberts-Williams:1994} C.D. Roberts and A.G. Williams,
Prog. Part. Nucl. Phys. {\bf 33}, 477 (1994).

\bibitem{Kizilersu:2009} A. Kizilersu and M.R. Pennington, Phys. Rev. D {\bf 79}, 125020 (2009).

\bibitem{Calcaneo:2011} A. Bashir, C. Calcaneo-Roldan, L.X. Gutiérrez-Guerrero and M.E.
Tejeda-Yeomans, Phys. Rev. D {\bf 83}, 033003 (2011).


\bibitem{Bashir:1998}
Z. Dong, H.J. Munczek and C.D. Roberts, Phys. Lett. B {\bf 33}, 536 (1994);
A. Bashir, A. Kizilersu and M.R. Pennington, Phys. Rev. D {\bf 57} 1242 (1998);

\bibitem{Bashir:2000} A. Kizilersu, M. Reenders and M.R. Pennington, Phys. Rev. D {\bf 52},
1242 (1995); A. Bashir, A. Kizilersu and M.R. Pennington, Phys.
Rev. D {\bf 62}, 085002 (2000); e-Print: hep-ph/9907418 [hep-th]
(1999); A. Bashir and A. Raya, Phys. Rev. D {\bf 64}, 105001
(2001).

\bibitem{Ward:1950} J.C. Ward, Phys. Rev. {\bf 78}, 1 (1950); H.S. Green, Proc. Phys. Soc. (London)
A {\bf 66}, 873 (1953; Y. Takahashi, Nuovo Cimento {\bf 6}, 371 (1957).

\bibitem{Bashir-1:1994} A. Bashir and M.R. Pennington, Phys. Rev. D {\bf 50}, 7679 (1994).

\bibitem{Bashir-2:1996} A. Bashir and M.R. Pennington, Phys. Rev. D {\bf 53}, 4694 (1996).

\bibitem{Landau:1956} L.D. Landau and I.M. Khalatnikov, Zh. Eksp. Teor. Fiz. {\bf 29}, 89
(1956); Sov. Phys. JETP {\bf 2}, 69 (1956); E.S. Fradkin,
Sov. Phys. JETP {\bf 2}, 361 (1956); K.Johnson and B. Zumino, Phys. Rev. Lett. {\bf 3} 351
(1959).

\bibitem{Bashir:LKF-1} A. Bashir, Phys. Lett. B {\bf 491}, 280 (2000); A. Bashir and A. Raya,
Phys. Rev. D {\bf 66} 105005 (2002).

\bibitem{Craig:2009} L. Chang and C.D. Roberts, Phys. Rev. Lett. {\bf 103} 081601 (2009).





\bibitem{Ball:1980} J.S. Ball and T-W. Chiu, Phys. Rev. D {\bf 22} 2542 (1980).







\bibitem{Roberts-Maris:1997} P. Maris and C.D. Roberts,
Phys. Rev. C {\bf 56}, 3369 (1997).

\bibitem{Tandy-Maris:2000} P. Maris and P.C. Tandy,
Phys. Rev. C {\bf 62}, 055204 (2000).

\bibitem{Tandy-Maris:2002} P. Maris and P.C. Tandy,
Phys. Rev. C {\bf 65}, 045211 (2002).

\bibitem{Ji-Maris:2001} C-R. Ji and P. Maris,
Phys. Rev. D {\bf 64}, 014032 (2001).

\bibitem{Tandy-Maris:1999} P. Maris and P.C. Tandy,
Phys. Rev. C {\bf 60}, 055214 (1999).

\bibitem{Tandy-Maris:2003} D. Jarecke, P. Maris and P.C. Tandy
Phys. Rev. C {\bf 67}, 035202 (2003).

\bibitem{Maris:2002} P. Maris,
PiN Newslett. {\bf 16}, 213 (2002).

\bibitem{Cotanch-Maris:2002} S.R. Cotanch and P. Maris,
Phys. Rev. D {\bf 66} 116010 (2002).

\bibitem{Bashir-Guerrero-1:2010} L.X.
Guti\'errez-Guerrero, A. Bashir, I.C. Cl\"oet and C.D. Roberts,
Phys. Rev. C {\bf 81}, 065202 (2010).

\bibitem{Bashir-Guerrero-2:2010} H.L.L. Roberts, C.D. Roberts, A. Bashir,
L.X. Guti\'errez-Guerrero, P.C. Tandy,
Phys. Rev. C {\bf 82}, 065202 (2010).

\bibitem{Tandy-Nguyen:2011} T. Nguyen, A. Bashir, C.D. Roberts, P.C. Tandy,
Phys. Rev. C {\bf 83}, 062201 (2011).




\bibitem{Roberts-Chang:2009} L. Chang and C.D. Roberts,
Phys. Rev. Lett. {\bf 103}, 081601 (2009).


\bibitem{ReviewCraig:2012} {\em "Collective perspective on advances in
Dyson-Schwinger Equation QCD"}, A. Bashir, L. Chang, I.C. Cloet,
B. El-Bennich, Y-X. Liu, C.D. Roberts and P.C. Tandy, Commun.
Theor. Phys. {\bf 58}, 79 (2012).

\bibitem{Skullerud:2003} J-I. Skullerud, P.O. Bowman, A. Kizilersu,
D.B. Leinweber and A.G. Williams
JHEP {\bf 04}, 047 (2003).

\bibitem{Skullerud:2007} A. Kizilersu, D.B. Leinweber, J-I. Skullerud and A.G. Williams,
Eur. Phys. J. C {\bf 50}, 871 (2007).

\bibitem{Bashir:LKF-2} A. Bashir and R. Delbourgo, J. of. Phys. A {\bf 37}, 6587 (2004);
A. Bashir and A. Raya, Nucl. Phys. B {\bf 709}, 307 (2005); Few
Body Syst. {\bf 41}, 185 (2007); A. Bashir and A. Raya, AIP Conf.
Proc. {\bf 1026}, 115 (2008); A. Bashir and A. Raya, J. Phys.
Conf. Ser. {\bf 37}, 90 (2006).

\bibitem{Bashir:2009} A. Bashir, A. Raya, S. S\'anchez-Madrigal and C.D. Roberts, Few Body
Sys. {\bf 46}, 229 (2009).

\bibitem{gluon} P. O. Bowman et al., Phys. Rev. D {\bf 76}, 094505
(2007); {\em "Unquenching the gluon propagator with
Schwinger-Dyson equations"}, A.C. Aguilar, D. Binosi, J.
Papavassiliou, arXiv:1204.3868 [hep-ph], (2012); {\em "Quark
flavour effects on gluon and ghost propagators"}, A. Ayala, A.
Bashir, D. Binosi, J. Rodr\'{\i}guez-Quintero, e-Print:
arXiv:1208.0795 [hep-ph] (2012).

\bibitem{Yamawaki:1989} K.-I. Kondo, H. Mino and K. Yamawaki,
Phys. Rev. D {\bf 39}, 2430 (1990); W.A. Bardeen, S.T. Love and
V.A. Miransky, Phys. Rev. D {\bf 42} 3514 (1990).

\bibitem{Rakow:1991} P.E.L. Rakow, Nucl. Phys. B {\bf 356}, 27
(1991).

\bibitem{Kondo:1990} K.-I. Kondo, Int. J. Mod. Phys. A {\bf 6}, 5447
(1990).
















\end{thebibliography}
\end{document}